\documentclass[english,aps,prb,twocolumn, superscriptaddress,showpacs,amssymb, amsmath]{revtex4-1}
\usepackage{graphicx}
\usepackage{float}
\usepackage{multirow}	
\usepackage{csquotes}	
\usepackage[table]{xcolor}
\usepackage[colorlinks=true,linkcolor=blue,citecolor=blue]{hyperref}
\usepackage{chemformula}
\usepackage{pdfpages}
\makeatletter
\AtBeginDocument{\let\LS@rot\@undefined}
\makeatother
\newcommand\marktopleft[1]{%
	\tikz[overlay,remember picture] 
	\node (marker-#1-a) at (0,1.5ex) {};%
}
\newcommand\markbottomright[1]{%
	\tikz[overlay,remember picture] 
	\node (marker-#1-b) at (0,0) {};%
	\tikz[overlay,remember picture,thick,dashed,inner sep=3pt]
	\node[draw,rounded rectangle,fit=(marker-#1-a.center) (marker-#1-b.center)] {};%
}
\usepackage{tikz}
\usetikzlibrary{fit,shapes.misc}
\usepackage{color, soul}	
\begin{document}
\title{Equilibration of Quantum hall edges in symmetry broken bilayer graphene}

\author{Chandan Kumar}
\author{Saurabh Kumar Srivastav}
\author{Anindya Das}
\email{anindya@iisc.ac.in}
\affiliation{Department of Physics, Indian Institute of Science, Bangalore 560 012, India}
\begin{abstract}
Equilibration of quantum Hall edges is studied in a high quality dual gated bilayer graphene device in both unipolar and bipolar regime when all the degeneracies of the zero energy Landau level are completely lifted. We find that in the unipolar regime when the filling factor under the top gate region is higher than the back gate filling factor, the equilibration is partial based on their spin polarization. 
 However, the complete mixing of the edge states is observed in the bipolar regime irrespective of their spin configurations due to the Landau level collapsing at the sharp pn junction in our thin hBN ($\sim$ 15 nm) encapsulated device, in consistent with the existing theory.

\end{abstract}


\maketitle
\section{Introduction}
Since the discovery of fractional quantum Hall (QH) effect in two-dimensional electron gas there have been extensive research on the equilibration of edge states to understand their properties\cite{kane1994randomness,kane1997quantized,bid2010observation}.
 Graphene, a single atomic layer of carbons, presents an unique platform where 
the equilibration of edge states along the pn junction gives rise to fractional values of conductance even in the integer QH regime\cite{williams2007quantum,ozyilmaz2007electronic,
liu2008fabrication,lohmann2009four,velasco2009electrical,ki2009quantum,ki2010dependence,
velasco2010probing,jing2010quantum,woszczyna2011graphene,nam2011ballistic,velasco2012quantum,bhat2012dual,
schmidt2013mixing,klimov2015edge,morikawa2015edge,kumada2015shot,matsuo2015edge,kumar2017equilibriation}. This is possible because the conduction and valance bands touch each other at the Dirac point, thus selective and control doping combined with the chiral nature of charge carriers can have co-propagating edge states along the pn junction interface. This co-propagating edge states has been used recently to demonstrate the mach zehnder interferometer in graphene\cite{wei2017mach}, where the selective equilibration between the symmetry broken QH edges determines the visibility of the interferometer. 

Equilibration of edge states have been studied extensively
\cite{williams2007quantum,ozyilmaz2007electronic,
	liu2008fabrication,lohmann2009four,velasco2009electrical,ki2009quantum,ki2010dependence,
	velasco2010probing,jing2010quantum,woszczyna2011graphene,nam2011ballistic,velasco2012quantum,bhat2012dual,
	schmidt2013mixing,klimov2015edge,morikawa2015edge,kumada2015shot,matsuo2015edge,kumar2017equilibriation} in graphene in both unipolar and bipolar regimes. The conductance plateaus observed are in agreement with the theoretical prediction by Abanin \textit{et al.}\cite{abanin2007quantized}. However, the conductance values in the unipolar regime for broken symmetry states in graphene device  deviates from the predicted equilibration values \cite{abanin2007quantized} and were found to be in good match with the partial equilibration based on spin polarized edge states\cite{amet2014selective}. Similar to graphene, the experiments are also performed on bilayer graphene (BLG), which has even more symmetries like orbital symmetry together with spin and valley symmetries. This leads to more complex Landau level (LL) phase diagram in BLG, which can be controlled independently by density, electric field and magnetic field\cite{lee2014chemical,kou2014electron,maher2014tunable,leroy2014emergent,hunt2017direct,zibrov2017tunable}.
The equilibration experiments performed in BLG\cite{jing2010quantum} also echo with the theory \cite{abanin2007quantized}.
However, no equilibration study has been performed on the broken symmetry states of ultra clean BLG devices either in unipolar or bipolar regime.

In this article, we report on the equilibration of QH edge states in a high quality dual gated bilayer graphene device in both unipolar (n-n$^*$-n/p-p$^*$-p) and bipolar (n-p-n/p-n-p) regime when all the degeneracy of the zero-energy LL is lifted. 
We find that in the unipolar regime when the top gate filling factor ($\nu_{TG}$) is higher than the back gate filling factor ($\nu_{BG}$), the conductance values does not follow the full equilibration prediction\cite{abanin2007quantized}.
Rather, they follow partial equilibration\cite{amet2014selective} based on the hierarchical splitting\cite{zibrov2017tunable,hunt2017direct} of zero energy LL with different spin configuration. 
Although the partial equilibration based on spin polarization is able to explain most of the conductance values in the unipolar regime but still it is unable to capture the conductance value for all the edge states. The lack of equilibration is better understood by considering the predominant mixing between the nearest edge states. Moreover, in the bipolar regime we find full equilibration of QH edge states for all combination of $\nu_{BG}$ and $\nu_{TG}$ irrespective of their spin configurations. The equilibration in the bipolar regime is understood in terms of LL
collapsing at the sharp pn junction as predicted by Lukose \textit{et al}.\cite{lukose2007novel}, NMR Peres \textit{et al}. \cite{peres2007algebraic}, Gu \textit{et al}.\cite{gu2011collapse} and LaGasse \textit{et al}.\cite{lagasse2016theory}.


	\begin{figure*}[t!]
		\includegraphics[width=0.8\textwidth]{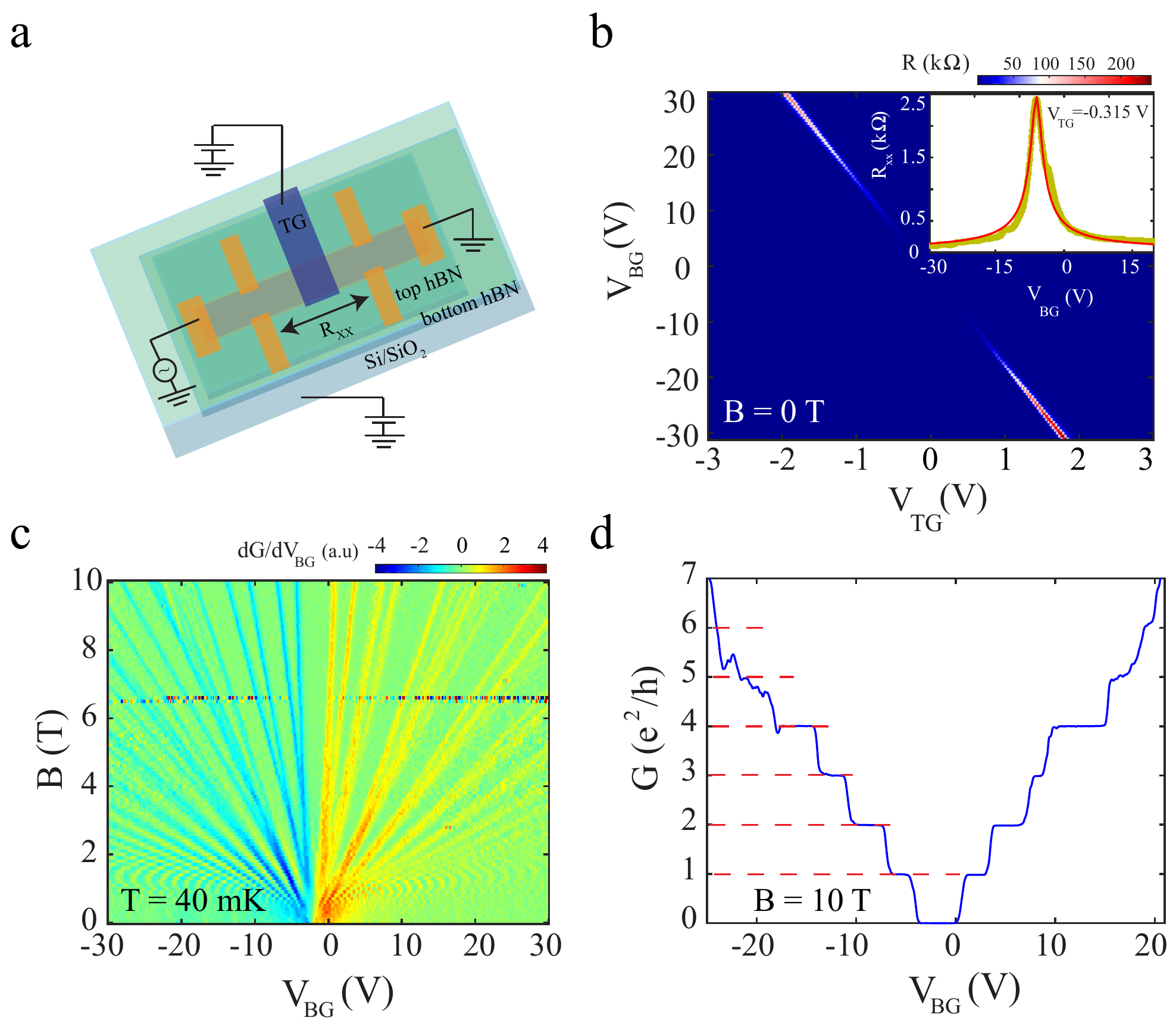}
		\caption{(a) The schematic of the encapsulated hBN/Bilayer graphene/hBN device. The 300 nm thick \ch{SiO2}  acts as back gate while the top thin hBN acts as a top gate and controls density only in middle portion of BLG. (b) 2D color plot of resistance as a function of back gate and top gate voltage at B=0 T. The inset shows the resistance at $V_{TG}=-0.315$ V. The red line is fit to resistance to extract mobility. (c) 2D color plot of transconductance $dG/dV_{BG}$ as a function of $V_{BG}$ and perpendicular magnetic field, $B$. Clear Landau levels can  be seen emitting from $V_{BG} = -1.3 $ V at B=0 T. (d) Two probe resistance as a function of back gate voltage at zero top gate voltage measured at $40$ mK and $10$ T magnetic field. All the degeneracy of the zero energy level is lifted, leading to the observation of QH plateaus at integer multiple of $e^2/h$.\label{Fig:fig1}}		
	\end{figure*}

\section{Device Fabrication and Characterization}

The bilayer graphene device is prepared using the dry transfer pick up technique\cite{wang2013one} using following steps. First, a glass slide is prepared with a layer of pdms and ppc. This glass slide is used to pick up the desired hBN ($\sim 10-15$ nm) flake, which is exfoliated on silicon wafer. On a separate silicon wafer graphite flakes are exfoliated and a bilayer graphene flake is picked up using the glass slide containing the pdms, ppc and hBN. The glass slide containing the heterostructure of pdms/ppc/hBN/BLG is then transfered on a thick hBN ($\sim \text{30 nm}$) which was already exfoliated on a separate silicon wafer. With this technique graphene remains in its pristine form as it is not exposed to any environmental contaminants or the pmma. The prepared stack of hBN/BLG/hBN is then cleaned in chloroform followed by acetone and IPA cleaning. Using the standard lithography technique the contacts are fabricated followed by etching in \ch{CHF3} and \ch{O2} environment. The etch rate is optimized to be 30 $\text{nm}$ per minute. Soon after etching Cr(3 nm)/Pd(8 nm)/Au(70 nm) is deposited at a base pressure of 3e-7 mbar. The top gate is fabricated by doing another lithography on the prepared stack of hBN/Gr/hBN. 
The top gate acts as a local gate and controls the density only in the middle portion of the device while the thick \ch{SiO2}  acts as a global back gate, controlling the density throughout the device as shown schematically in Fig. \ref{Fig:fig1}a, where the contacts are shown in yellow color and top gate is shown in blue color. The device is 5 $\mu m$ long, 2 $\mu m$ wide and the separation between inner contacts is 2.5 $\mu m$. 

The conductance is measured using standard lock in technique. The different combination of back gate ($V_{BG}$) and top gate ($V_{TG}$) voltages leads to the formation of n-p-n/p-n-p or n-n$^*$-n/p-p$^*$-p region in the same device as shown in Fig. \ref{Fig:fig1}b. The diagonal line in Fig. \ref{Fig:fig1}b corresponds to the Dirac point under the top gated region, where the density is controlled by both $V_{BG}$ and $V_{TG}$. From the slope of diagonal line we calculate top hBN thickness of $\sim 18$ nm. The inset shows resistance measured as a function of $V_{BG}$ for $V_{TG} = -0.315$ V and from the fitting we obtain a mobility of $40,000$ $cm^2/Vs$. Figure \ref{Fig:fig1}c displays landau level fan diagram ($dG/dV_{BG}$) as a function of  $V_{BG}$ and magnetic field (B) at $V_{TG}=0$ V. 
Figure \ref{Fig:fig1}d shows the two probe conductance at $B=10$ T, where one can clearly see the QH plateaus at integer multiple of $e^2/h$ suggesting the lifting of spin, valley and orbital degeneracy of the zero energy LL of BLG.

\section{Equilibration of QH edges}

\begin{figure}[t!]
	\includegraphics[width=.48\textwidth]{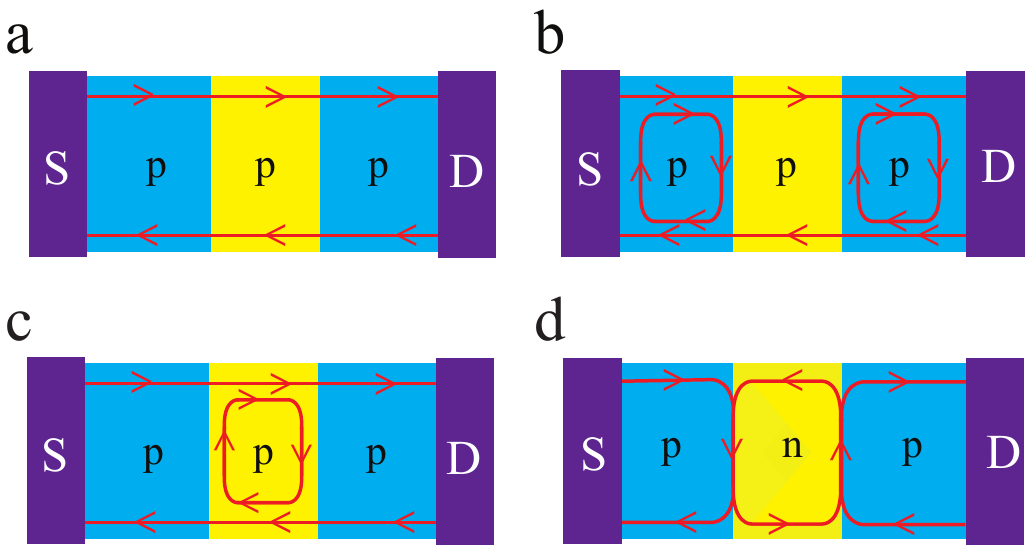}
	\caption{(a) Schematic of the chiral QH edge state (a) For $\nu_{BG}=\nu_{TG}=-1$, here since $|\nu_{BG}|=|\nu_{TG}|$, all the $\nu_{BG}$ edge states are completely transmitted through the top gate region. 
	(b) For $|\nu_{BG}|>|\nu_{TG}|$, in this scenario the only $|\nu_{TG}|$ edge states are completely transmitted and the rest ($|\nu_{BG}|-|\nu_{TG}|$) edge states gets reflected back. (c) For $|\nu_{BG}|<|\nu_{TG}|$, all the $|\nu_{BG}|$ edge states completely transmitted through top gate region while the $|\nu_{TG}|-|\nu_{BG}|$ edge states keep circulating in the top gate region. These inner circulating edge states can now interact with the transmitting edge states and the conductance gets modified depending on the equilibration. (d) Since electron and hole edge states have opposite chirality, the electron and hole edge states move co-propagating along the pn junction and the full equilibration of these edge states leads to value of conductance given by eqn. \ref{eqn_bipolar}.\label{Fig:fig2}}
\end{figure}

Two probe conductance in the QH regime depends on the back gate ($\nu_{BG}$)  and top gate ($\nu_{TG}$) filling factors. This is shown schematically in Fig. \ref{Fig:fig2}. The unipolar regime (i.e top gate and back gate region has same kind of charge carrier) is shown in Fig. \ref{Fig:fig2}a-c. When $\nu_{BG} = \nu_{TG}$, then the current injected from the back gate region completely transmits through the top gate region without any back scattering. For $|\nu_{BG}| > |\nu_{TG}|$, the extra edge states in the back gate region ( $|\nu_{BG}| - |\nu_{TG}|$) gets reflected back as shown in Fig. \ref{Fig:fig2}b. Hence, the conductance is determined only by the number of edge states under the top gate region and thus, the total conductance is given by
\begin{equation}
G_{pp^*p/nn^*n}= min (\left|\nu_{BG}\right|,\left|\nu_{TG}\right|)\frac{e^2}{h}
\label{eqn_unipolar_same}
\end{equation}
More interesting situation arises when $|\nu_{TG}| > |\nu_{BG}|$. In this case $|\nu_{BG}|$ edge channels are completely transmitted through the top gate region while $|\nu_{TG}| - |\nu_{BG}|$ number of edges keep circulating under the top gate region. The $|\nu_{BG}|$ edge states transmitting through the top gate region can mix with the circulating $|\nu_{TG} - \nu_{BG}|$ channels under the top gate region, which leads to the modification of two probe conductance. In the case of complete mixing, which has been observed \cite{williams2007quantum,ozyilmaz2007electronic,
	liu2008fabrication,lohmann2009four,velasco2009electrical,ki2009quantum,ki2010dependence,
	velasco2010probing,jing2010quantum,woszczyna2011graphene,nam2011ballistic,velasco2012quantum,bhat2012dual,
	schmidt2013mixing,klimov2015edge,morikawa2015edge,kumada2015shot,matsuo2015edge,kumar2017equilibriation} on \ch{SiO2} 
substrate and the conductance is given as:
\begin{equation}
G_{pp*p/nn*n}= \frac{\left|\nu_{BG}\right| \left|\nu_{TG}\right|}{2 \left|\nu_{TG}\right|- \left|\nu_{BG}\right|}\frac{e^2}{h}
\label{eqn_unipolar}
\end{equation}
Another interesting scenario arises if the charge carrier in the top gate and the back gate region are of different type; holes and electrons. Since electron and hole has opposite chirality, the net two probe conductance will be zero unless the clock wise moving QH edge states can equilibrate with the anti-clock wise moving edge states along the pn junction interface as shown in Fig. \ref{Fig:fig2}d. For the complete equilibration case the two probe conductance is given as:
\begin{equation}
G_{pnp/npn}= \frac{\left|\nu_{BG}\right| \left|\nu_{TG}\right|} { 2 \left| \nu_{TG}\right|+ \left| \nu_{BG}\right|}\frac{e^2}{h}
\label{eqn_bipolar}
\end{equation}

 The conductance values discussed so far corresponds to the cases when the degeneracy like spin, valley or orbital are not lifted. Recent experiment\cite{amet2014selective} by Amet \textit{et al}. studies the equilibration (unipolar regime) in graphene  when both the valley and spin degeneracies are completely lifted. In such scenario they have showed that the equilibration depends strongly on the spin configuration of the edge states. If the transmitting edge states ($\nu_{BG}$) and inner circulating edge states under the top gate region (Fig. \ref{Fig:fig2}c) have opposite spin then the edge state do not equilibrate and hence the conductance is given by $\nu_{BG} \times e^2/h $, on the contrary if the edge states have same spin, they equilibrate completely and conductance is given by Eq. \ref{eqn_unipolar}. Thus, the conductance for spin polarized edge state in the unipolar regime can be given as\cite{amet2014selective}:
\begin{equation}
G_{\rm{partial}}=\sum_{i=\downarrow,\uparrow}\frac{\left|\nu_{TG,i}\right|\left|\nu_{BG,i}\right|}{2\left|\nu_{TG,i}\right|-\left|\nu_{BG,i}\right|} \frac{e^2}{h}
\label{eqn_partial}
\end{equation}
Here, $ \nu_{TG,\uparrow} (\nu_{TG,\downarrow})$ refers to the total number of edge states with $\uparrow (\downarrow)$, same convention holds for $\nu_{BG}$.
For example, if $\nu_{BG}=1$ with $ \nu_{BG,\uparrow}=1$ and $\nu_{TG}=2$ with $ \nu_{TG,\uparrow}=1$ and $\nu_{TG,\downarrow}=1$, then $ G= 2e^2/3h$. On the other hand, if $\nu_{TG}=2$ with $ \nu_{TG,\downarrow}=2$ and $ \nu_{TG,\uparrow}=0$ 
then $G=0$.

	\begin{figure}[ht!]
	\includegraphics[width=0.45\textwidth]{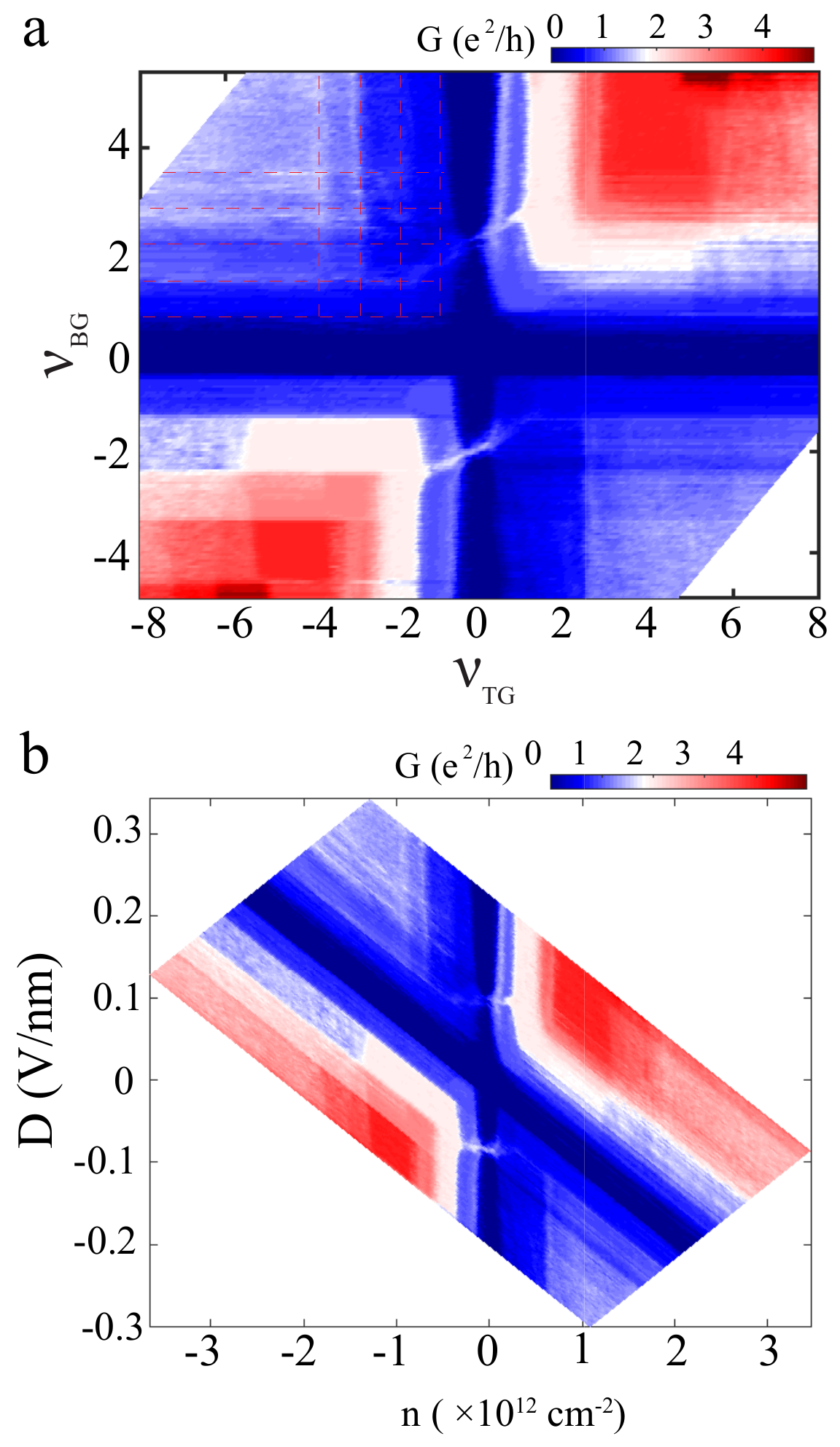}
	\caption{(a) 2D plot of conductance as a function of $\nu_{BG}$ and $\nu_{TG}$ at 10 T magnetic field. The horizontal and vertical strips in the unipolar regime corresponds to back gate and top gate filling factor, receptively. Quantized conductance of $2e^2/h$ is observed for $\nu_{BG}=-2$ from $\nu_{TG}$=-2 to $\nu_{TG}$=-5. Similarly for  $\nu_{BG}=-4$, quantized conductance of $4e^2/h$ is observed from $\nu_{TG}$=-4 to $\nu_{TG}$=-5. In the bipolar region we observe full equilibration of edge states. The chequerboard pattern is represented by red dashed in bipolar regime. (b) 2D plot of displacement field as a function of electron density. The LL crossing point is at $D \sim 0.9 \text{V/nm}$.  \label{Fig:fig3}}	
\end{figure}

\begin{figure*}[t]
	\centering
	\includegraphics[width=1.0\textwidth]{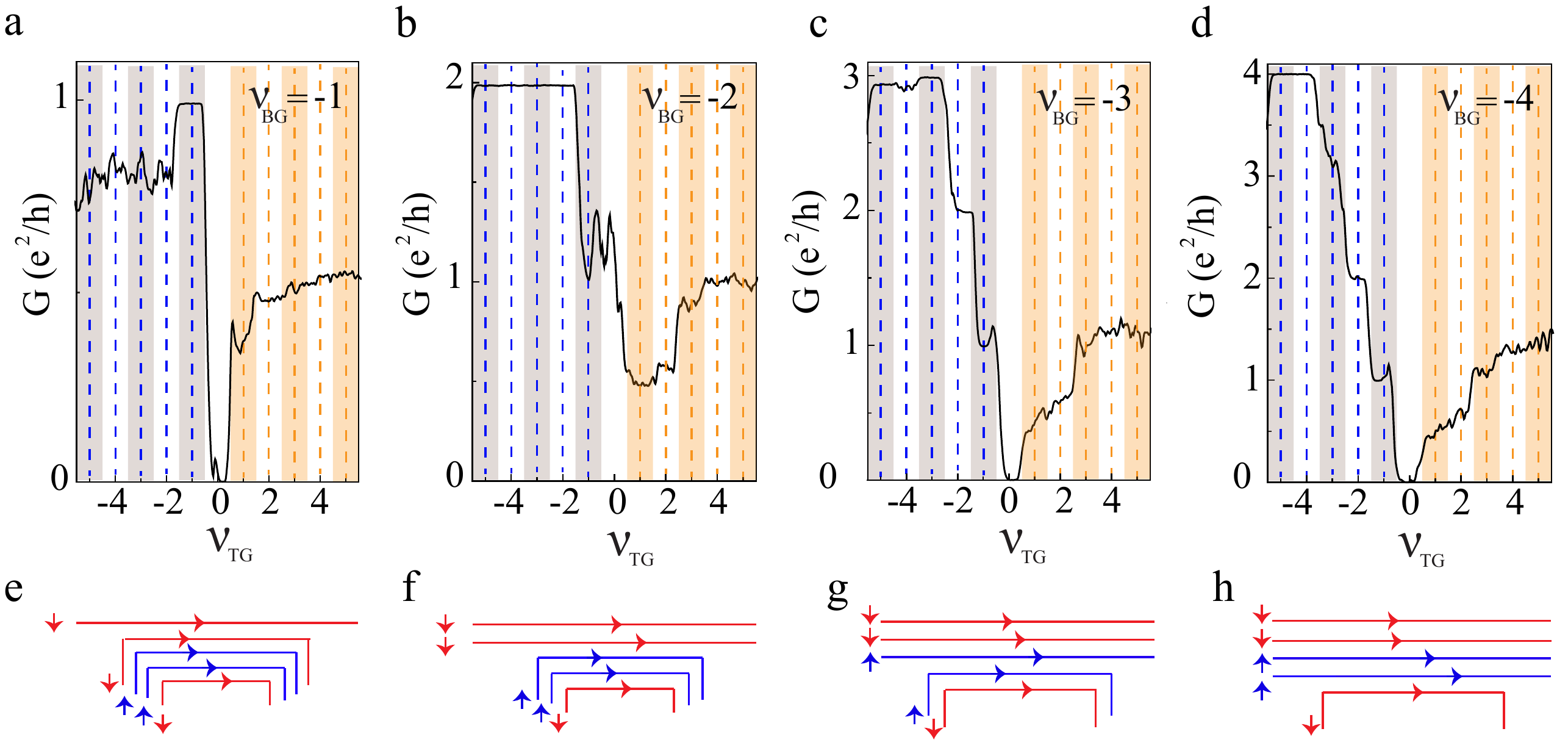}
	\caption{Line trace of G as a function of $\nu_{TG}$ at different set of $\nu_{BG}$ at 10 T and 40 mK. (a) Cut across $\nu_{BG}=-1$. (b) Cut across $\nu_{BG}=-2$. (c) Cut across $\nu_{BG}=-3$. (d) Cut across $\nu_{BG}=-4$. For  $\nu_{BG}=-2$ ($\nu_{BG}=-4$)  conductance remains quantized at $2e^2/h$ ( $4e^2/h$) from  $\nu_{TG}=-2$ to $\nu_{TG}=-5$ ($\nu_{TG}=-4$ to $\nu_{TG}=-5$), suggesting that these edge states do not equilibrate in our device. The each figure in the lower panel represents the different spin configuration associated with each of the LL. The spin states are obtained from Fig. \ref{Fig:fig5}a below \lq b \rq. The up and down spin are represented by blue and red color, respectively. \label{Fig:fig4}}
	
\end{figure*}

\section{Results}

The color plot of two probe conductance as a function of $\nu_{TG}$ and $\nu_{BG}$ at 10 T is shown in Fig. \ref{Fig:fig3}a. The Fig. \ref{Fig:fig3}a is obtained from Fig. 1 of the SI, where the conductance is measured as a function of $V_{BG}$ and $V_{TG}$ at 10 T. The back gate and top gate voltages is converted to back gate and top gate filling factors, respectively, using $\nu= \frac{nh}{4eB}$, where $n$ is the density in the system, $h$ is the planks constant, $e$ is the electronic charge and $B$ is the applied magnetic field. The horizontal and vertical strips in Fig. \ref{Fig:fig3}a corresponds to different $\nu_{BG}$ and $\nu_{TG}$, respectively. Interestingly, in the unipolar regime, for $\nu_{BG} = -2$ conductance remains constant at a value of $2e^2/h$ from $\nu_{TG}= -2$ to $\nu_{TG}= -5$. We also observe a similar feature for $\nu_{BG} = -4$, where quantized conductance of $4e^2/h$ is observed from $\nu_{TG}= -4$ to $\nu_{TG}= -5$. On the other hand in the bipolar regime we observe a clear chequerboard pattern with the conductance values (Table I) expected for the full equilibration case (Eq. \ref{eqn_bipolar}). Two lines of higher conductance are also observed connecting the unipolar and bipolar region. Note that these lines are placed symmetrically about $\nu_{BG} = 0$. Fig. \ref{Fig:fig3}b shows color plot of conductance as a function of density ($n$) and displacement field ($D$). The net density ($n$) and the displacement field ($D$) are obtained using following relation\cite{zhang2009direct}:
$n = (D_B-D_T)/e$ and $ D=(D_B+D_T)/2$.
Here $D_B$ and $D_T$ are the applied back gate and top gate displacement
field, receptively. They are given as
 $ D_T=\frac{\epsilon_t(V_{TG}-V_{TG}^{th})}{d_t}$ and $ D_B=\frac{\epsilon_b(V_{BG}-V_{BG}^{th})}{d_b}$ where ($d_t,d_b$) are the thickness of dielectric layers, ($\epsilon_t, \epsilon_b$) are the dielectric constants and ($V_{TG}^{th},V_{BG}^{th}$) is the charge neutrality points (CNP) of the device. 
 The high conductance line joining $\nu_{TG}=-1$ and $\nu_{TG}=1$ LLs \cite{lee2014chemical,kou2014electron,maher2014tunable,leroy2014emergent,hunt2017direct,zibrov2017tunable} is at $D \sim 0.9 \text{V/nm}$.\\
 
 

 Figure \ref{Fig:fig4}a-d plots the cut lines obtained from Fig. \ref{Fig:fig3}a at different $\nu_{BG}$. In the unipolar regime expected quantized plateaus are observed for $\nu_{TG}=\nu_{BG}$. For $|\nu_{TG}| < |\nu_{BG}|$ also the conductance plateaus agrees with Eq. \ref{eqn_unipolar_same} as can be seen in Fig. \ref{Fig:fig4}c and \ref{Fig:fig4}d. However, For $|\nu_{TG}| > |\nu_{BG}|$ the conductance plateaus is not consistent with full equilibration prediction as mentioned in Eq. \ref{eqn_unipolar}. In-fact for $\nu_{BG} = -2$ the conductance plateau remains at $2e^2/h$ from $\nu_{TG}=-2$ to $\nu_{TG}=-5$. Similarly, Fig. \ref{Fig:fig4}d shows a plateau of $4e^2/h$ from $\nu_{TG}=-4$ to $\nu_{TG}=-5$ for $\nu_{BG}=-4$. For the full equilibration of QH edge states the conductance values of $3/2,4/3,5/4$ for $\nu_{TG}=-3,-4,-5$ was expected for $\nu_{BG}=-2$ and a value of $3.33 e^2/h$ for $\nu_{BG},\nu_{TG}= -4,-5$. This gives a very clear signature that some of the edge states, in particular $\nu_{BG}=-2$ prefer not to equilibrate with the circulating edges under the top gate region. Even for $\nu_{BG}= -1$ and $\nu_{BG}= -3$ (Fig. \ref{Fig:fig4}a and Fig. \ref{Fig:fig4}c) the conductance values obtained in the unipolar regime is higher than the full equilibration prediction. Similar behavior is also observed in the four probe resistance measurement(SI). 
 However, in the bipolar region full equilibration of edge states are observed.
 Although the data in the bipolar regime do not show good plateau as observed in our previous work on single layer graphene\cite{kumar2017equilibriation} but its average value matches with Eq. \ref{eqn_bipolar}. The measured conductance values in both unipolar and bipolar regimes have been compared with the theoretical prediction based on full equilibration shown in Table I. Each round bracket in Table I, from left to right lists the conductance value obtained using  full equilibration followed by experimental value. The large mismatch of conductance observed in the unipolar regime between the theory and experiment is highlighted by dashed square.


\begin{figure}[t!]
	\centering
	\includegraphics[width=0.5\textwidth]{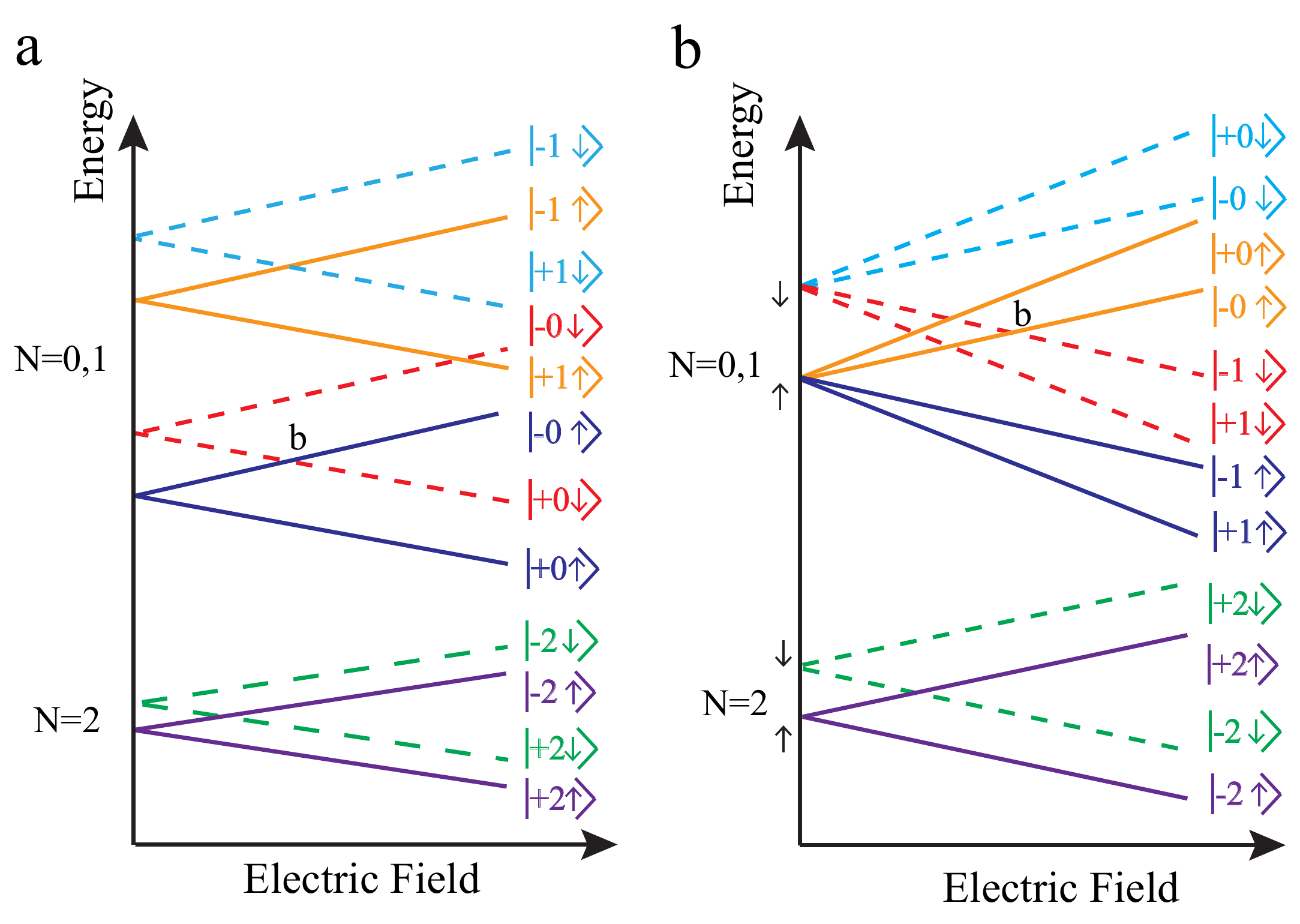}
	\caption{Schematic representation of LL evolution with magnetic field and the perpendicular electric field. The solid (dashed) line represents the up(down) spin states while the orbital index and valley is represented by (0,1) and (+,-) respectively. (a) Model adapted from ref \cite{hunt2017direct,zibrov2017tunable}. (b) Model adapted from ref \cite{lee2014chemical,kou2014electron,maher2014tunable,leroy2014emergent}.\label{Fig:fig5}}
\end{figure}


\begin{table*}[!htb]
	\caption{Conductance values based on full equilibration in Unipolar and Bipolar regime (theory vs experiment)}
	\begin{minipage}{.5\linewidth}
		\centering
				\begin{tabular}{lccccc}
					\hline\hline \\[.1ex] 
					$\nu_{BG} \backslash \nu_{TG}$ & -1 & -2 & -3 & -4 & -5\\ [1ex] 
					\hline\\[.1ex]
					-1 &  (1; 1) & \marktopleft{c2}(0.67; 0.8) & (0.6; 0.8) & (0.57; 0.8) &  (0.55; 0.8)\markbottomright{c2}\\[2ex]
					-2 &  (1; 1.1) & (2; 2) &\marktopleft{c2} (1.5; 2) & (1.33; 2) & (1.25; 2)\markbottomright{c2}\\[2ex]
					-3 &  (1; 1) & (2; 2) & (3; 3) & \marktopleft{c2}(2.4; 2.92) & (2.14; 2.93)\markbottomright{c2}\\[2ex]
				    -4 &  (1; 1) & (2; 2) & (3; 3.1) & (4; 4) &\marktopleft{c2}(3.33; 4)\markbottomright{c2} \\[2ex] 
					\hline 
					\hline
				\end{tabular}
	\end{minipage}%
	\begin{minipage}{.56\linewidth}
		\centering
		\begin{tabular}{ccccc}
			\hline\hline \\[.1ex] 
			 1 & 2 & 3 & 4 & 5\\ [1ex] 
			 \hline\\[.1ex]
		      (0.33; 0.32) & (0.4; 0.49) & (0.43; 0.51) & (0.44; 0.54) &  (0.45; 0.56)\\[2ex]
			  (0.5; 0.51) & (0.66; 0.58) & (0.75; 0.83) & (0.8; 0.96) &  (0.83; 0.98)\\[2ex]
			  (0.6; 0.44) & (0.86; 0.74) & (1; 1.1) & (1.09; 1.2) & (1.15; 1.26)\\[2ex]
		      (0.67; 0.56) & (1; 0.79) & (1.2; 1.12) & (1.33; 1.3) & (1.43; 1.39) \\[2ex] 
			\hline 
			\hline
		\end{tabular}
	\end{minipage} 
\end{table*}

%
\section{Discussions}

In the following section we try to understand the equilibration data in unipolar regime. The equilibration of QH edge state in unipolar regime can be understood by considering the hierarchical splitting of the zero energy LL\cite{lee2014chemical,kou2014electron,maher2014tunable,leroy2014emergent,hunt2017direct,zibrov2017tunable}.


The Landu levels (LL) energy in BLG is given by \cite{mccann2006landau,novoselov2006unconventional} $E_n = \pm \hbar \omega \sqrt{N(N-1)} $, where $\omega$ is the cyclotron frequency, $\omega = eB/m^*$ and $m^*\sim 0.033 m_e$ is the effective mass in BLG and $N$ is a non-negative integer representing LL index in each layer. For each orbital number, $N$, each of the LL is four fold degenerate due to two valley and two spin degeneracy.
Thus, the zero energy ($N=0,1$) LL in BLG is eight fold degenerate\cite{mccann2006landau,guinea2006electronic}.
Hence, the sequential splitting of the zero energy LL is very intricate. Various efforts have been made to understand the hierarchical splitting of zero energy LL in BLG\cite{lee2014chemical,kou2014electron,maher2014tunable,leroy2014emergent,hunt2017direct,zibrov2017tunable}.\\ 
Figure \ref{Fig:fig5}a shows the model by Zibrov \textit{et al.}\cite{zibrov2017tunable} and Hunt \textit{et al.}\cite{hunt2017direct}, in this model at finite magnetic field 
both the orbital and spin degeneracy is lifted and the application of electric field lifts the valley degeneracy\cite{hunt2017direct,zibrov2017tunable}. The other model is shown in Fig.\ref{Fig:fig5}b, here the finite magnetic field 
only lifts the spin degeneracy and the electric field is responsible for lifting the valley and orbital degeneracy \cite{lee2014chemical,kou2014electron,maher2014tunable,leroy2014emergent}.\\
\begin{table}[h!]
	\begin{center}
		\caption{Conductance values based on partial equilibration in Unipolar Regime (theory vs experiment)}
		\begin{tabular}{lccccc}
			\hline\hline \\[.2ex] 
			$\nu_{BG} \backslash \nu_{TG}$ & -1 & -2 & -3 & -4 & -5\\ [2ex] 
			\hline\\[.1ex]
			-1 &  (1; 1) &  \hl{(0.67; 0.8)} & \hl{(0.67; 0.8)} &\hl{(0.67; 0.8)} &  \hl{(0.60; 0.8)}\\[2ex]
			-2 &  (1; 1) & (2; 2) & (2; 2) & (2; 2) &  \hl{(1.5; 2)}\\[2ex]
			-3 &  (1; 1) & (2; 2) & (3; 3) &\hl{(2.67; 2.92)} &\hl{(2.17; 2.93)} \\[2ex]
			-4 &  (1; 1) & (2; 2) & (3; 3) & (4; 4) & \hl{(3.5; 4)} \\[2ex] 
			\hline 
			\hline
		\end{tabular}
	\end{center}	
\end{table}
We find that our data can be explained by model presented in Fig. \ref{Fig:fig5}a [ Fig. \ref{Fig:fig5}b] if the $D$ (electric field) lies below [above] the LL crossing value marked by \enquote{b} in Fig. \ref{Fig:fig5}a [Fig. \ref{Fig:fig5}b]. 
In that scenario the different spin configuration associated with each LL edge sates 
is shown in the bottom row of Fig.\ref{Fig:fig4}. The red and blue color denote the up and down spin, respectively. Table II lists the conductance value obtained using partial equilibration (Eqn. \ref{eqn_partial}) and the experimental data. Each round bracket in Table II, from left to right lists the conductance value obtained using  partial equilibration (Eqn. \ref{eqn_partial}) followed by experimental value. We would like to mention that our data is more suitable with Fig. \ref{Fig:fig5}a\cite{zibrov2017tunable,hunt2017direct} as can be seen from Fig. \ref{Fig:fig3}b that the values of $D$ for the unipolar edges in Table I is far below the $D = 0.9 \text{V/nm}$, at which LL crossing happens (\enquote{b} marker in Fig. \ref{Fig:fig5}a).


We find that although partial equilibration model\cite{amet2014selective} can explain data for $\nu_{BG}=-2$ from $\nu_{TG}=-2$ to $\nu_{TG}=-4$ but it fails to explain the quantized conductance value at $\nu_{BG, TG}= -2,-5$ and $\nu_{BG, TG}= -4,-5$.
Moreover, the conductance value obtained at different $\nu_{TG}$ for $\nu_{BG}=-1$ and $\nu_{BG}=-3$ are also not consistent with the partial equilibration model. The inconsistency between the values obtained by partial equilibration model and the experimental data is highlighted with yellow color in Table II, where one can clearly see that experimental values are always higher than the partial equilibration model suggesting lack of equilibration between the transmitting edges and the circulating edges under the top gate region.\\

The partial equilibration model assume the equal amount of equilibration between the transmitting edges ($\nu_{BG}$) and the circulating edges ($\nu_{TG}-\nu_{BG}$) under the top gate region depending on their spin polarization irrespective of their spacial location (bottom panel of Fig. \ref{Fig:fig4}). The lack of equilibration in our experiment suggests that the equal amount of equilibration between all the edges may not be completely true and possibly the    
equilibration between the nearest edge states of transmitting edge ($\nu_{BG}$) and circulating edge ($|\nu_{TG}|=|\nu_{BG}|+1$) in Fig. \ref{Fig:fig4} is stronger compared to innermost circulating edge. 
For example, when $\nu_{BG}=-2$ the conductance value will be $2e^2/h$ for $\nu_{TG}= -3$ and $-4$ due to opposite spin configuration, however, for $\nu_{TG}= -5$ the equilibration between $\nu_{BG}=-2$ and $\nu_{TG}= -5$ (same spin polarization) will be very weak as they are spatially separated as well as due to the screening by the inner $\nu_{TG}= -3$ and $-4$ edges in Fig. \ref{Fig:fig4}. Similarly, it can explain for Fig. \ref{Fig:fig4}c-\ref{Fig:fig4}g and and Fig. \ref{Fig:fig4}d-\ref{Fig:fig4}h. However, further theoretical studies are required to understand the equilibration for symmetry broken edges in bilayer graphene.

We now focus in the bipolar regime where the complete mixing of edge states is observed (Table I) eventhough the spin polarizations of the edges are opposite as can be seen in Fig. \ref{Fig:fig5}. However, this is not surprising as LaGasse \textit{et al.}\cite{lagasse2016theory} have shown that for the $\sim 40$ nm width of p-n junction 
 the landau levels are superimposed on each other as it is comparable to magnetic length scale and the full mixing of edge states are observed. In our top gated device geometry we estimate the width of pn junction to be $\sim 15-20$ nm, and thus the edges at the pn junction are not well defined. Not only that in the bipolar regime strong in-plane electric field is present across our sharp p-n junction and studies \cite{lukose2007novel,gu2011collapse,peres2007algebraic} predict the collapse of LL due to the effective higher magnetic length scale. However, further studies are required to understand the exact nature of edge mixing mechanism in BLG.\\

\section{Conclusion}
In conclusion, our study provides the first experimental evidence of edge state equilibration in bilayer graphene when all the degeneracies of zeroth energy level are completely lifted. Although the partial equilibration based on spin polarization is able to explain most of the conductance values in the unipolar regime but still unable to capture for all the edge states. The lack of equilibration is better understood by considering the predominant mixing between the nearest edges. In the bipolar regime full equilibration is observed irrespective of the spin configurations and understood in terms of LL collapsing at the sharp pn junction.

\bibliographystyle{apsrev4-1}
\bibliography{ref}
\onecolumngrid
\newpage
\thispagestyle{empty}
\mbox{}
\includepdf[pages=-]{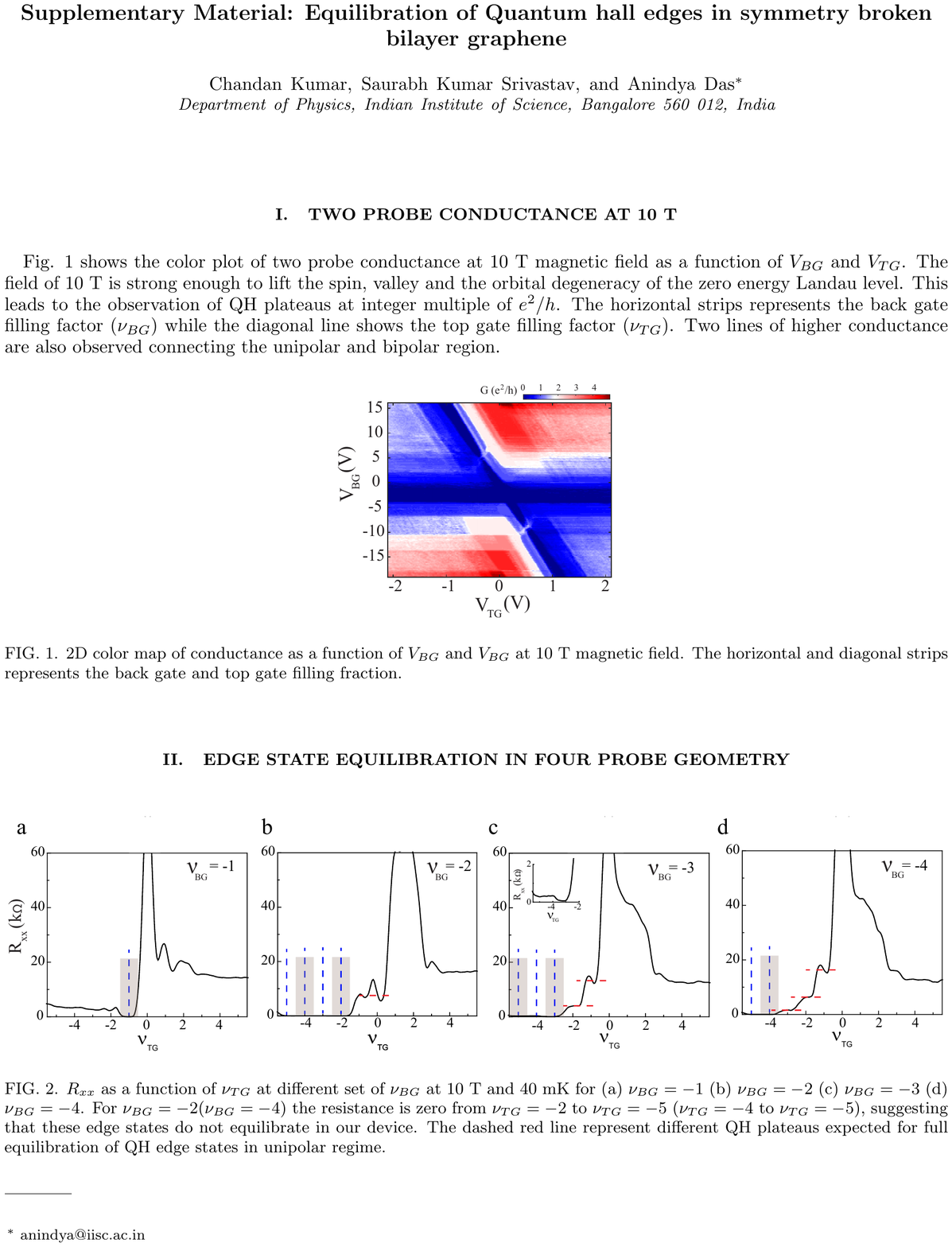}
\end{document}